\documentstyle[12pt]{article}

\catcode`@=11

\def\beqra{\begin{eqnarray}} \def\eeqra{\end{eqnarray}}
\def\beqast{\begin{eqnarray*}} \def\eeqast{\end{eqnarray*}}
\def\beq{\begin{equation}}      \def\eeq{\end{equation}}
\def\be{\begin{enumerate}}   \def\ee{\end{enumerate}}

\def\fo{\hbox{{1}\kern-.25em\hbox{l}}}
\def\fnote#1#2{\begingroup\def\thefootnote{#1}\footnote{#2}\addtocounter
{footnote}{-1}\endgroup}

\def\sppt{Research supported in part by the
Robert A. Welch Foundation and NSF Grant PHY 9009850}

\def\utgp{Theory Group\\ Department of Physics \\ University of Texas
\\ Austin, Texas 78712}

\def\ul{\underline}

\def\ch{\@startsection{section}{1}{\z@}{-3ex plus-1ex minus-.2ex}%
        {2ex plus.2ex}{\large\sc}}

\def\raisenot{\raise .5mm\hbox{/}}
\def\nota{\ \hbox{{$a$}\kern-.49em\hbox{/}}}
\def\notA{\hbox{{$A$}\kern-.54em\hbox{\raisenot}}}
\def\notb{\ \hbox{{$b$}\kern-.47em\hbox{/}}}
\def\notB{\ \hbox{{$B$}\kern-.60em\hbox{\raisenot}}}
\def\notc{\ \hbox{{$c$}\kern-.45em\hbox{/}}}
\def\notd{\ \hbox{{$d$}\kern-.53em\hbox{/}}}
\def\notbd{\ \hbox{{$D$}\kern-.61em\hbox{\raisenot}}} 
\def\note{\ \hbox{{$e$}\kern-.47em\hbox{/}}}
\def\notk{\ \hbox{{$k$}\kern-.51em\hbox{/}}}
\def\notp{\ \hbox{{$p$}\kern-.43em\hbox{/}}}
\def\notq{\ \hbox{{$q$}\kern-.47em\hbox{/}}}
\def\notW{\ \hbox{{$W$}\kern-.75em\hbox{\raisenot}}}
\def\notz{\ \hbox{{$Z$}\kern-.61em\hbox{\raisenot}}}

\def\notpa{\hbox{{$\partial$}\kern-.54em\hbox{\raisenot}}}

\def\7#1#2{\mathop{\null#2}\limits^{#1}}        
\def\5#1#2{\mathop{\null#2}\limits_{#1}}        

\def\inbar{\vrule height1.5ex width.4pt depth0pt}
\def\IB{\relax{\rm I\kern-.18em B}}
\def\IC{\relax\leavevmode\hbox{\,$\inbar\kern-.3em{\rm C}$}}
\def\ID{\relax{\rm I\kern-.18em D}}
\def\IE{\relax{\rm I\kern-.18em E}}
\def\IF{\relax{\rm I\kern-.18em F}}
\def\IG{\relax\leavevmode\hbox{\,$\inbar\kern-.3em{\rm G}$}}
\def\IH{\relax{\rm I\kern-.18em H}}
\def\II{\relax{\rm I\kern-.18em I}}
\def\IK{\relax{\rm I\kern-.18em K}}
\def\IL{\relax{\rm I\kern-.18em L}}
\def\IM{\relax{\rm I\kern-.18em M}}
\def\IN{\relax{\rm I\kern-.18em N}}
\def\IO{\relax\leavevmode\hbox{\,$\inbar\kern-.3em{\rm O}$}}
\def\IP{\relax{\rm I\kern-.18em P}}
\def\IQ{\relax\leavevmode\hbox{\,$\inbar\kern-.3em{\rm Q}$}}
\def\IR{\relax{\rm I\kern-.18em R}}
\def\sed{\hbox{{\sf S}\kern-.4em\hbox{\sf S}}}
\def\ZZ{\relax{\sf Z\kern-.4em Z}}
\def\smIR{\hbox{{\footnotesize\rm I}\kern-.2em\hbox{\footnotesize\rm R}}}
\def\smIO{\ \hbox{{\footnotesize\rm I}\kern-.4em\hbox{\footnotesize\bf O}}}
\def\smIQ{\ \hbox{{\footnotesize\rm I}\kern-.5em\hbox{\footnotesize\bf Q}}}
\def\IGa{\relax{\rm I}\kern-.18em\Gamma}
\def\IPi{\relax{\rm I}\kern-.18em\Pi}
\def\IQt{\relax\leavevmode\hbox{$\kern.3em\inbar\kern-.3em\Theta$}}
\def\IOm{\relax\hbox{$\kern3.48pt\inbar\kern1.8pt\inbar\kern-5.28pt\Omega$}}

\def\ca#1{\relax\ifmmode {\cal#1} \else$\cal#1$\fi}     
\def\Sf#1{\relax\ifmmode\hbox{\sf#1}\else{\sf#1}\fi}    
\def\fibby{\ifcase\@ptsize                      
                \font\tenrm=cmfib8\or           
                \font\elvrm=cmfib8 scaled\magstephalf\or        
                \font\twlrm=cmfib8 scaled\magstep1 \fi}         
\def\TeXey{\ifcase\@ptsize\or\or                
                \font\twlrm=cmr10 scaled\magstep1       
                \font\twlmi=cmmi10 scaled\magstep1      
                \font\twlit=cmti10 scaled\magstep1      
                \font\twlbf=cmbx10 scaled\magstep1\fi}  

\def\ch{\@startsection{section}{1}{\z@}{-3ex plus-1ex minus-.2ex}%
        {2ex plus.2ex}{\large\sc}}
\def\sch{\@startsection{subsection}{2}{\z@}{-1.5ex plus-1ex minus-.2ex}%
        {1pt plus.2ex}{\sc}}
\def\ssch{\@startsection{subsubsection}{3}{\z@}{-1ex plus-1ex minus-.2ex}%
        {1pt plus.2ex}{\small\sc}}
\def\seceq{\@addtoreset{equation}{section}
        \def\theequation{\thesection.\arabic{equation}}}        

\def\con{\ifmmode \hbox{\bf*} \else{\bf*}\fi}   
\def\scon{\ifmmode \hbox{\footnotesize\rm\bf*} \else{\footnotesize\rm\bf*}\fi}

\def\0#1{\relax\ifmmode\mathaccent"7017{#1}
        \else\accent23#1\relax\fi}              

\def\place#1#2#3{\vbox to0pt{\kern-\parskip\kern-7pt
                             \kern-#2truein\hbox{\kern#1truein #3}
                             \vss}\nointerlineskip}

\def\illustration #1 by #2 (#3){\vbox to #2{\hrule width #1 height 0pt depth
0pt
                                       \vfill\special{illustration #3}}}

\def\scaledillustration #1 by #2 (#3 scaled #4){{\dimen0=#1 \dimen1=#2
           \divide\dimen0 by 1000 \multiply\dimen0 by #4
            \divide\dimen1 by 1000 \multiply\dimen1 by #4
            \illustration \dimen0 by \dimen1 (#3 scaled #4)}}

\catcode`@=12

\begin{document}

\hfill{UTTG-36-91}

\hfill{December 1991}

\vfill
\vspace{24pt}
\begin{center}

{\bf Light-Cone Gauge for 1+1 Strings\fnote{*}{\sppt}}

\vspace{24pt}

Eric Smith

\vspace{12pt}
\utgp
\vspace{36pt}

\ul{ABSTRACT}
\end{center}
\baselineskip=24pt

Explicit construction of the light-cone gauge quantum theory of
bosonic strings in 1+1 spacetime dimensions reveals unexpected
structures.  One is the existence of a gauge choice that gives a free
action at the price of propagating ghosts and a nontrivial BRST
charge.  Fixing this gauge leaves a U(1) Kac-Moody algebra of
residual symmetry, generated by a conformal tensor of rank two and
a conformal scalar.  Another is that the BRST charge made from these
currents is nilpotent when the action includes a linear dilaton
background, independent of the particular value of the dilaton
gradient.  Spacetime Lorentz invariance in this theory is still
elusive, however, because of the linear dilaton background and the
nature of the gauge symmetries.

\vfill
\pagebreak
\setcounter{page}{1}

\ch{Introduction}

The invention of matrix models\cite{EHM} opened a whole new pathway to
the understanding of quantum string theories and even string field
theories.  However, since the domain of matrix models is limited to
dimensions less than one, the problem of inventing higher
dimensional (presumably more physically realistic) string field
theories is still approachable only in the continuum formulation.  The
hope of applying what has been learned in less than one dimension to
problems stated in higher dimensions thus focuses attention on the
correspondence between matrix models and continuum theories in the one
case for which both are defined, strings in one dimension.

The one-dimensional noncritical theory of strings actually has two
degrees of freedom and may be cast as a critical theory of strings in
$1 + 1$ dimensions in the presence of nontrivial background
fields.  Understanding of the discreet-continuum correspondence would
benefit greatly from an explicit continuum formulation of first a
quantum theory and then a field theory of strings in $1 + 1$
dimensions.  The construction of the quantum theory will be undertaken
here.

One issue that must be addressed in quantization is the choice of
gauge in which to work.  One usually thinks of a choice
between conformal and light-cone gauges as a choice
between manifest covariance at the price of nontrivial
ghosts and a ghost-free spectrum at the price of hidden covariance.
However, because one-dimensional string theory has no transverse
dimensions, there is another relevant consideration in this problem,
the possibility of obtaining a free action for the dynamical fields
which remain after gauge fixing.  In string field theory,
manifest covariance actually becomes an inconvenience, making it
tricky to consistently define the string interaction
vertices\cite{Zweibach}, so the intended string field application of
this work recommends a light-cone gauge, in which the string vertices
are simply specified\cite{CnG}.  Further, the presence of a nontrivial
dilaton background in the critical $1+1$ theory makes it possible to choose
an off-diagonal world-sheet/spacetime light-cone gauge which gives a free
action.  This choice, however, leaves a residual reparametrization
symmetry which leads to propagating ghosts and a nontrivial BRST charge.

We will use this gauge, and sacrifice both manifest covariance and
ghost-triviality in exchange for a free action and a simpler vertex in
the future field theory.  By checking the nilpotency of the BRST
charge, we will ensure that the reparametrization symmetry we want
is not violated by anomalies.  The outline of this work is as follows.

Section 2 introduces the assumptions that define the theory.
Quantization will be done in the language of functional integrals,
in which anomalies are seen as violations by the
measure of the gauge symmetries which have been built into the action.
In this work, the choices of defining symmetry and the appropriate forms
for the action and measure to implement them will be drawn from
previous work done with the Liouville theory and other linear
dilaton backgrounds.

Section 3 contains the treatment of the Faddeev-Popov gauge-fixing of
the theory and the resulting well-defined functional integral.
The BRST charge is constructed in section 4, and related to
the generators of residual symmetry in the gauge-fixed functional
integral.  Finally section 5 points out some of the nonstandard
features that arise in this construction.

\ch{String Theories in Two Dimensions}

We wish to build a quantum theory of strings in two
spacetime dimensions, where possible using forms that will make
second quantization easier.  We specify the theory
by its desired symmetries, and express physical amplitudes as
correlation functions of the form
\beq
<\cdots> = \int {\cal D} fields \, {e}^{iS} \cdots
     					\label{eq:corrfn}
\eeq
where \( \cdots \) is some function of the fields invariant under
the desired symmetry group.

As usual, we introduce a world sheet with metric, on which
the spacetime embedding coordinates appear as scalar fields.  Bosonic
string theory in other than $26$ dimensions can be cast in two forms,
either as a noncritical theory in which world sheet diffeomorphism is
a symmetry but variation of the world sheet geometry is not, or as a
critical theory with one more dimension, in which both diffeomorphism
and geometric variation of the world sheet are symmetries.
The form of the action in the critical version of the
theory is\cite{David,DnK,Joe}
\beq
{S}_{C} = {\frac{-1}{4 \pi \alpha ' }} \int {d}^{2} \sigma \sqrt{g}
\left(
{g}^{ab} {\partial}_{a}{X}^{\mu} {\partial}_{b}{X}^{\nu}
         {\eta}_{\mu \nu}
+ \alpha ' n\cdot X {R}^{(2)} \right)
     						\label{eq:caction}
\eeq
where \( \alpha ' \) is the inverse string tension,
\( {g}_{ab} \) is the world sheet metric, and
\( {X}^{\mu} \) are the fields that embed the string in spacetime.
\( {\eta }_{\mu \nu } \) are the components of the flat spacetime metric,
\( n \) is a constant spacetime vector and the term linear in \( X \)
is the dilaton background.

World sheet diffeomorphism invariance of~(\ref{eq:caction}) is automatic,
and the pseudo-Weyl variation
\beq
{g}_{ab} \rightarrow (1 + \epsilon) {g}_{ab} ; \; \; \;
{X}^{\mu} \rightarrow {X}^{\mu} - \frac{\alpha ' {n}^{\mu} }{2} \epsilon .
     						\label{eq:pseudoweyl}
\eeq
will also be a symmetry of the quantum theory if the term linear in
${R}^{(2)}$ in the resulting variation of the action is scaled correctly
to cancel the variation of the measure\cite{Polyakov}.  The scaling
expected from the Liouville theory is the same as that
proposed by Myers in a slightly different context\cite{Myers}.  Considering
general linear dilaton backgrounds, he has shown that if the
generators of Lorentz
transformations of the embedding fields are supplemented with
generators which transform \( n \), the algebra of the Lorentz
group can be recovered if the norm
\( {n}^{2} = \frac{26 - D}{6\alpha '} \), where \( D \) is the
dimension of the embedding spacetime.  Thus \( n \) is spacelike when
\( D < 26 \) and when \( D = 26 \) the
dilaton gradient vanishes and one obtains the standard simplest
critical bosonic string action.  We will work in \( D = 2 \) and
include the nonzero dilaton gradient, but will allow it to take some
general spacelike norm.

We will find that the gauge choice that makes the
action~(\ref{eq:caction}) free results in a nontrivial Faddeev-Popov
determinant and so leads to propagating ghosts.  The naive version of
this gauge gives dynamics to both reparametrization and Weyl ghosts in
an unnecessarily complicated way.  This can be avoided if we work with
the Weyl-invariant fields
\beq
{\hat{g}}^{ab} \equiv \sqrt{g} {g}^{ab}; \; \; \;
{\hat{X}}^{\mu} \equiv {X}^{\mu} +
                       \frac{\alpha ' {n}^{\mu} }{2} \log{\sqrt{g}}
     						\label{eq:weylfield}
\eeq
in terms of which the action becomes
\begin{eqnarray}
\lefteqn{{S}_{C} = {\frac{-1}{4 \pi \alpha ' }} \int {d}^{2} \sigma
\left[ {\hat{g}}^{ab} {\partial}_{a}{\hat{X}}^{\mu}
                      {\partial}_{b}{\hat{X}}^{\nu}
                      {\eta}_{\mu \nu}
           + \alpha ' n\cdot \hat{X} {\hat{R}}^{(2)}
\vphantom{ - \frac{ {\alpha '}^{2} {n}^{2} }{2} \log{\sqrt{g} }
\left({\hat{R}}^{(2)} - \frac{1}{2} \widehat{\Box} \log{\sqrt{g} }
\right) }\right. }
                                                      \nonumber \\
& &  \phantom{ = {\frac{-1}{4 \pi \alpha ' }} \int {d}^{2} \sigma }
\left. \mbox{} - \frac{ {\alpha '}^{2} {n}^{2} }{2} \log{\sqrt{g} }
\left({\hat{R}}^{(2)} - \frac{1}{2} \widehat{\Box} \log{\sqrt{g} }
\right) \right]
     						\label{eq:wcaction}
\end{eqnarray}
where \( {\hat{R}}^{(2)} \) and \( \widehat{\Box} \) are respectively
the scalar curvature and the covariant Laplacian in the ``metric''
\( \hat{g} \).  The natural light-cone gauge to make the
action~(\ref{eq:wcaction}) free gives a simpler ghost action, from which
the Weyl ghost can be eliminated by an equation of constraint, leaving
all residual symmetry in reparametrizations.  It would be nice to find
a gauge which eliminates propagating ghosts altogether and still gives
a free action, but that has not been accomplished here.  The
action~(\ref{eq:wcaction}) will be the starting point in our construction of
the quantum theory.

\ch{Fixing Light-Cone Gauge}

The world sheet time and space coordinates are labeled
\( \left( \tau ,\sigma \right) \)
and the spacetime embedding fields are \( {X}^{+},{X}^{-} \), in terms of
which the spacetime metric components are
${\eta}_{+-} = {\eta}_{-+} = -1$,
${\eta}_{++} = {\eta}_{--} = 0$.  In
a light-cone gauge one of the world sheet
coordinates is specified in terms of one of the spacetime embedding
fields, such as \( {X}^{+} = \tau \).  In addition, two other degrees
of freedom of the world sheet metric must be constrained in
order to fix the gauge freedom.  To do this it is convenient to introduce
world sheet coordinates \( {\sigma}^{\pm} \equiv \frac{1}{\sqrt{2}}
\left( \tau \pm \sigma \right) \).

It was emphasized by Tzani\cite{Tzani} that the world sheet metric cannot
be fixed in the conformal gauge once light-cone coordinates are
chosen because that would overly constrain the system.  Therefore, in
this gauge one of the metric components must remain dynamical.
Polyakov\cite{PKPZ} has pointed out that an off-diagonal gauge, such as
\( {g}^{--} = 0, \: \sqrt{g} = 1 \) produces a simple measure for the
remaining degree of freedom of the metric, which can be regulated by
modifying the propagator without introducing vertices.  In this theory
it will also produce the desired free action for the remaining fields.
Therefore we will make the gauge choice
\begin{eqnarray}
{\hat{X}}^{+}  & = & \tau \nonumber \\
{\hat{g}}^{--} & = & 0
     						\label{eq:gchoice} \\
\sqrt{g}       & = & 1,        \nonumber
\end{eqnarray}
It will be convenient for later steps to define a scale
\( \frac{1}{m} \equiv \frac{\alpha ' {n}^{+}}{\sqrt{2}} \).

General diffeomorphisms and Weyl transformations are parametrized by
vector field components \( {\epsilon }^{+}, \, {\epsilon}^{-} \) and a
Weyl parameter \( {\epsilon } \).  The infinitessimally shifted fields
are given by
\begin{eqnarray}
{\hat{X}}^{+\epsilon }  & = & {\hat{X}}^{+}
                        - {\epsilon }^{a} {{\hat{X}}^{+},}_{a}
                        - \frac{1}{ \sqrt{2} m } { {\epsilon }^{a},}_{a}
                          \nonumber \\
{\hat{g}}^{--\epsilon } & = & {\hat{g}}^{--}
                             - {({\epsilon }^{a}{\hat{g}}^{--}),}_{a}
                             + 2 {{\epsilon }^{-},}_{-}{\hat{g}}^{--}
                             + 2 {{\epsilon }^{-},}_{+}{\hat{g}}^{+-}
     						\label{eq:efield}\\
{\sqrt{g}}^{\epsilon } & = & \sqrt{g}
                             - {({\epsilon }^{a}{\sqrt{g}}),}_{a}
                             + \epsilon \sqrt{g} \nonumber
\end{eqnarray}
In the Faddeev-Popov gauge-fixing procedure\cite{FnP} a well defined
and gauge-fixed functional integral is given by
\beq
Z = \int {\cal D }\left(
         {\hat{g}}^{ab},{\sqrt{g}},{\hat{X}}^{\mu}\right)
\delta \left[ \sqrt{2}\left( {\hat{X}}^{+} - \tau \right) \right]
\delta \left[ {\hat{g}}^{--} \right]
\delta \left[ \sqrt{g} - 1 \right]
{\Delta}_{FP}
{e}^{i{S}_{C}}
     						\label{eq:gfcorrfn}
\eeq
where the Faddeev-Popov determinant is given in terms of the variation
of the gauge conditions with respect to the generators of the gauge
symmetry:
\beq
{\Delta}_{FP} \equiv
{
\left. \det
\left[ \frac{\partial
\left( \sqrt{2}{\hat{X}}^{+\epsilon },
               {\hat{g}}^{--\epsilon },
               {\sqrt{g}}^{\epsilon } \right) }{\partial
\left( {\epsilon }^{+},
       {\epsilon }^{-},
       {\epsilon } \right) } \right]
\right|
}_{ {\epsilon }^{+},{\epsilon }^{-},{\epsilon } = 0 }
     						\label{eq:deltafp}
\eeq
Adding sources coupled to additional terms in the action
in~(\ref{eq:gfcorrfn}) produces the
desired generator of the correlation functions~(\ref{eq:corrfn}).
Physically meaningful correlation functions are generated by sources
that couple to field quantities invariant under diffeomorphisms
and the transformation~(\ref{eq:pseudoweyl}).

\ch{Residual Symmetry and the BRST Charge}

One way to expose the final group of invariances of the
integral~(\ref{eq:gfcorrfn})
is to study its BRST symmetry\cite{BRST}.  By introducing auxiliary fields
\( {B}_{f}, \, B, \, {\hat{B}}_{--} \), antighosts
\( {b}_{f}, \, b, \, {\hat{b}}_{--} \) and ghosts
\( {c}^{+}, \, {c}^{-}, \, c \), one can write the same integral as
\beq
Z = \int {\cal D }\left(
         {\hat{g}}^{ab},{\sqrt{g}},{\hat{X}}^{\mu},
         {B}_{f},{B},{\hat{B}}_{--},
         {b}_{f},{b},{\hat{b}}_{--},
         {c}^{+},{c}^{-},{c}\right)
{e}^{i( {S}_{C}+{S}_{G.F.}+{S}_{GH} )},
     						\label{eq:qcorrfn}
\eeq
where
\begin{eqnarray}
{S}_{G.F.} & = & \int {d}^{2}\sigma
                 \left[ {B}_{f}
                        \sqrt{2} \left( {\hat{X}}^{+}-\tau \right)
                        + {\hat{B}}_{--} {\hat{g}}^{--}
                        + B\left( \sqrt{g} - 1 \right)
                 \right]
     						\label{eq:gfaction}\\
{S}_{GH}   & = & \int {d}^{2}\sigma
\begin{array}[t]{l}
  \left[ -{b}_{f}\left( \sqrt{2}{{\hat{X}}^{+},}_{a}{c}^{a}
                      + \frac{1}{m} {{c}^{a},}_{a} \right) \right. \\
              + {\hat{b}}_{--}\left( {{c}^{-},}_{-} {\hat{g}}^{--}
                                     - {c}^{-} {{\hat{g}}^{--},}_{-}
                                     - {({c}^{+}{\hat{g}}^{--}),}_{+}
                                     + 2 {{c}^{-},}_{+}{\hat{g}}^{+-}
                              \right)                             \\
  \left. -b\sqrt{g}\left( {{c}^{a},}_{a}
                               + {c}^{a}{\partial }_{a}\log \sqrt{g}
                               - c \right) \right]
\end{array}  					\label{eq:ghaction}
\end{eqnarray}

In the Faddeev-Popov procedure, one usually {\em discovers\/} a global
nilpotent symmetry of the classical action (for both matter and ghosts)
in~(\ref{eq:qcorrfn}),  called the
BRST symmetry.  Requiring that this transformation preserve its
classical properties in the full quantum theory is then the check
that anomalies
do not spoil the gauge invariance of the theory assumed at the
classical level.  This condition constrains the central charge of
the theory, which may appear as a restriction on the number of
embedding dimensions or in some other way.  In this string theory, the
gauge transformation~(\ref{eq:pseudoweyl}) is not a symmetry
of the action at the
classical level (the term from variation of the measure is required to
cancel the resulting term from variation of the action), so the
distinction between classical and quantum seems artificial.  However,
we can still guess what should be the form of the BRST transformation
on fields at the classical level from the algebra of
the gauge constraints, as one would do in the Kugo-Uehara
procedure\cite{KugUeha} for gauge fixing the operator Hamiltonian of a
constrained system.

The BRST symmetry is conveniently introduced using
an anticommuting parameter \( \theta \) to produce the following field
variations:
\beq
\begin{array}{l}
  \begin{array}{ll}
    \begin{array}{lcl}
    \delta {b}_{f}        & = & -\theta {B}_{f}        \\
    \delta {\hat{b}}_{--} & = & -\theta {\hat{B}}_{--} \\
    \delta b              & = & -\theta B
    \end{array}
  &
    \begin{array}{lcl}
    \delta {B}_{f}        & = & 0 \\
    \delta {\hat{B}}_{--} & = & 0 \\
    \delta B              & = & 0
    \end{array}
  \end{array}
\\
  \begin{array}{ll}
  \delta {c}^{a} = - \left( \theta {c}^{b} \right) {{c}^{a},}_{b} \; \; \;
  &
  \delta c       = - \left( \theta {c}^{b} \right) {c,}_{b}
  \end{array}
\\
  \begin{array}{lcl}
  \delta {\hat{X}}^{+}  & = & - \left( \theta {c}^{a} \right)
                                {{\hat{X}}^{+},}_{a}
                              - \frac{1}{\sqrt{2}m}
                               {\left( \theta {c}^{a} \right) ,}_{a}
  \\
  \delta {\hat{g}}^{ab} & = &  {\left( \theta {c}^{a} \right) ,}_{c}
                                {\hat{g}}^{cb}
                              +{\left( \theta {c}^{b} \right) ,}_{c}
                                {\hat{g}}^{ac}
                              -{\left( \theta {c}^{c} {\hat{g}}^{ab}
                                \right) ,}_{c}
  \\
  \delta \sqrt {g}      & = & -{\left( \theta {c}^{a} \sqrt{g}
                                \right) ,}_{a}
                              +\left( \theta c \right) \sqrt{g}
  \end{array}
\end{array}
     						\label{eq:qdelta}
\eeq
When \( \theta \) is a constant \( \delta \) should be a global symmetry
of the theory.  Here the term \( \theta \) has been written
as it would appear in general variations, such as one would make to
find the associated Noether current.
As well as being an expansion parameter, \( \theta \) is a
sign-tracking parameter, and when it is constant it is possible to use
it without listing it explicitly to define the unparametrized global
transformation \( \Delta \) by setting
\( \delta \equiv \theta \Delta \).
It is straightforward to check that \( {\Delta }^{2} = 0 \) as a
transformation of fields at the classical level.

The action of the BRST symmetry on the canonical variables is
precisely the action of the gauge symmetry, with the ghosts
substituted as the parameters in the gauge transformations, and
further it may be seen that the gauge-fixing and ghost actions
themselves come from a BRST variation,
\beq
{S}_{G.F.} + {S}_{GH} = \Delta \int {d}^{2} \sigma
\left[ -{b}_{f}\sqrt{2}( {\hat{X}}^{+} - \tau )
       -{\hat{b}}_{--} {\hat{g}}^{--}
       -b(\sqrt{g} - 1) \right]
     						\label{eq:kugo}
\eeq
Thus if the action and measure for the canonical fields respect the
gauge symmetries used to define $\Delta $, nilpotency ensures that it
it will be a global symmetry of the gauge fixing and ghost actions as well.
Therefore, it will have an associated conserved charge, called the BRST
charge (usually labeled $Q$), which
as an operator in the quantum theory is the generator of the BRST
symmetry.

A standard result\cite{BRST,Teitelboim,GnR} is that maintaining
nilpotency of the BRST variation in the quantum theory and identifying
physical states with the elements of the BRST cohomology class is
equivalent to preserving the physical gauge
equivalence relation in the quantum correlation functions.  Thus
the requirement that the classical nilpotency of the BRST
transformation be preserved as the operator statement
${Q}^{2} \equiv 0$ is our check that quantization preserves
the gauge symmetry we have assumed to define the theory.

While it is possible to calculate \( Q \) directly as a
Noether charge and compute its square using an operator product
expansion, the algebra of the residual symmetry left after gauge
fixing and its relation to
the original gauge invariance imposed on the theory can be more easily
seen if we express \( Q \) in terms of the components of the
energy-momentum tensor which generate the residual symmetry
transformations in the gauge-fixed functional
integral~(\ref{eq:qcorrfn})

To find these we first perform the integrations over the auxiliary
fields and the antighost \( b \) that couples to \( \sqrt{g} \).  The
auxiliary field integrations simply restore the original gauge-fixing
\( \delta \)-functionals, and because of the simple form of its ghost
term the \( b \) integration produces a \( \delta \)-functional which
constrains the value of the Weyl ghost \( c \).  Next integrating over
\( {\hat{g}}^{--}, \; \sqrt{g}, \; {\hat{X}}^{+} \) and \( c \)
places these fields on the constraint surface.  The resulting
functional integral is
\beq
Z = \int {\cal D }\left(
         {\hat{g}}^{++},{\hat{X}}^{-},
         {b}_{f},{\hat{b}}_{--},
         {c}^{+},{c}^{-}\right)
{e}^{i( {S}_{C}+{S}_{GH} )},
     						\label{eq:rgcorrfn}
\eeq
in which the remaining actions are reduced to
\begin{eqnarray}
{S}_{C}  & = & \frac{\sqrt{2}}{4 \pi \alpha '} \int {d}^{2} \sigma
               \left( 1 + \frac{{\partial }_{+}}{m} \right)
               {\hat{g}}^{++} {{\hat{X}}^{-},}_{+}
     						\label{eq:gfcaction} \\
{S}_{GH} & = & \int {d}^{2}\sigma
\left[  -{b}_{f}\left( 1 + \frac{{\partial }_{a}}{m} \right) {c}^{a}
        + 2 {\hat{b}}_{--} {{c}^{-},}_{+} \right]
     						\label{eq:gfghaction}
\end{eqnarray}
As desired the Weyl ghost is present neither explicitly nor through
possible BRST variations of the remaining dynamical quantities.

By construction the ghost equations of motion are precisely the
equations satisfied by gauge parameters that preserve the gauge slice.
 Thus the EOM for the reparametrization ghosts tell us what part of
the diffeomorphism invariance was left unfixed by the gauge condition.
 Making the general decomposition
\beq
\begin{array}{lcl}
{c}^{+} & = & -({1 + \frac{{\partial }_{-}}{m}}) \tilde{u}
          + {e}^{-\sqrt{2}m\tau} \tilde{v} \\
{c}^{-} & = & \tilde{u}
\end{array}
\;
\begin{array}{lcl}
       {b}_{f} & = & -m {e}^{\sqrt{2}m\tau} \tilde{z} \\
{\hat{b}}_{--} & = & \frac{\tilde{w}}{2}
                   - \frac{1}{2m} {e}^{\sqrt{2}m\tau} {\tilde{z},}_{-}
\end{array}
     						\label{eq:ghdecomp}
\eeq
the ghost action becomes
\beq
{S}_{GH} = \int {d}^{2} \sigma
    \left( \tilde{w}{\tilde{u},}_{+}
         + \tilde{z}{\tilde{v},}_{+} \right)
     						\label{eq:gdaction}
\eeq
and we find that the residual symmetry unfixed by the gauge condition
is the collection of reparametrizations generated by the vector field
components
\beq
\begin{array}{lclclcl}
{\epsilon }^{+} & = & -({1 + \frac{{\partial }_{-}}{m}})
                        {u}^{\epsilon }
                      + {e}^{-\sqrt{2}m\tau} {v}^{\epsilon }
                & \; \; \; &
{{u}^{\epsilon },}_{+} & = & 0 \\
{\epsilon }^{-} & = & {u}^{\epsilon }
                & \; \; \; &
{{v}^{\epsilon },}_{+} & = & 0
\end{array}
     						\label{eq:gpreparm}
\eeq
One may check directly that the algebra of these classical
transformations is a $U(1)$ Kac-Moody algebra, which we will see again
as the algebra of their associated Noether currents.

Now the form of the action~(\ref{eq:gfcaction}) is very simple, but
the measure in~(\ref{eq:rgcorrfn}) is not.  The presence of the Weyl
anomaly tells us that it is metric-dependent, and in this gauge the
metric has a dynamical component.  Therefore, as in the work of
David\cite{David}, Distler and Kawai\cite{DnK}, we make the ansatz
that the measure in~(\ref{eq:rgcorrfn}) may be written as a
field-independent measure times a counterterm, as
\beq
{\cal D }\left(
         {\hat{g}}^{++},{\hat{X}}^{-},
         {b}_{f},{\hat{b}}_{--},
         {c}^{+},{c}^{-}\right) =
\tilde{\cal D } \left(
         {\hat{g}}^{++},{\hat{X}}^{-},
         {b}_{f},{\hat{b}}_{--},
         {c}^{+},{c}^{-}\right)
{e}^{i( {S}_{M} )}.
\label{eq:newmetric}
\eeq
We can choose the counterterm by requiring that the reparametrizations
generated by~(\ref{eq:gpreparm}), and thus the BRST transformation,
be classical symmetries of the resulting improved canonical action
${S}_{C} + {S}_{M}$\footnote{This simplification was proposed by
J. Polchinski}.  This leads to the conclusion that on this gauge
surface of constraint
\beq
{S}_{M} = \frac{1}{4\pi \alpha '} \int {d}^{2} \sigma
          {\hat{g}}^{++}.
     						\label{eq:improv}
\eeq
It then remains to show that the measure $\tilde{\cal D }$ is indeed
free from anomalies by demonstrating nilpotency of the BRST charge in
the quantum theory.

Associated with the symmetries generated by
\( {u}^{\epsilon } \left( {\sigma }^{-} \right) \)
and
\( {v}^{\epsilon } \left( {\sigma }^{-} \right) \)
are Noether currents \( {J}_{u} \) and
\( {J}_{v} \), obtained by varying with general
\( {u}^{\epsilon } \left( {\sigma }^{-} , {\sigma }^{+} \right) \) and
\( {v}^{\epsilon } \left( {\sigma }^{-} , {\sigma }^{+} \right) \),
which obey \( {{J}_{u},}_{+} = 0; \; {{J}_{v},}_{+} = 0 \) when the
dynamical fields are on shell.  The equations of motion from
\( {S}_{C} + {S}_{M} \) are satisfied by functions of the form
\begin{eqnarray}
{\hat{g}}^{++} & = & \tilde{f} + {e}^{-\sqrt{2}m\tau } \tilde{g}
     						\label{eq:cgdecomp} \\
{\hat{X}}^{-}  & = & \tilde{h} + {e}^{ \sqrt{2}m\tau } \tilde{j} - \tau
     						\label{eq:cxdecomp}
\end{eqnarray}
with
\beq
\begin{array}{lclclcl}
{\tilde{f},}_{+} & = & 0; & \; & {\tilde{g},}_{+} & = & 0 \\
{\tilde{h},}_{+} & = & 0; & \; & {\tilde{j},}_{+} & = & 0
\end{array}
     						\label{eq:cshell}
\eeq
(If we were working in suitably
defined complex world sheet coordinates the fields
\( \tilde{f},\, \tilde{g},\, \tilde{h},\, \tilde{j},\,
   \tilde{u},\, \tilde{v},\, \tilde{w},\, \tilde{z} \) would be analytic
functions).

It is straightforward to show that with the proper transformation
properties for the ghosts, the diffeomorphisms generated
by~(\ref{eq:gpreparm})
are also symmetries of \( {S}_{GH} \).
In order that the ghost action~(\ref{eq:ghaction}) be a scalar under
diffeomorphism it is necessary that \( {b}_{f} \) transform as a
scalar density and \( {\hat{b}}_{--} \) as a component of a rank-2
contravariant tensor (\({\hat{g}}^{--} \) is already the component of
a covariant tensor density).
Similarly \( {c}^{+}, \, {c}^{-} \) are components of a covariant
vector.

Using these reparametrization properties, one may find the currents
for canonical fields and ghosts separately in terms of analytic
variables.  Their algebra is simplest if they are written using the
rescaled or shifted fields
\beq
\begin{array}{lcl}
f & = &  \tilde{f} + 2 \\
g & = &  \tilde{g}     \\
h & = &  \frac{\sqrt{2}}{4\alpha '}\tilde{h} \\
j & = & -\frac{\sqrt{2}}{4\alpha '}\tilde{j}
\end{array}
\; \; \;
\begin{array}{lcl}
u & = & \tilde{u} \\
v & = & \tilde{v} \\
w & = & \pi \tilde{w} \\
z & = & \pi \tilde{z}
\end{array}
\label{eq:currentfields}
\eeq
The expression for the canonically normalized currents in terms of
these fields is
\beq
\begin{array}{lcl}
{{J}_{u}}_{C} & = &
\left( -f{h,}_{-} - {g,}_{-}j - 2g{j,}_{-}
       + \frac{2}{m}{h,}_{--} + \frac{1}{2\alpha '} \right) \\
{{J}_{v}}_{C} & = &
\left( mfj + 2{j,}_{-} \right) \\
{{J}_{u}}_{GH} & = &
\left( -v{z,}_{-} - {w,}_{-}u -2w{u,}_{-} \right) \\
{{J}_{v}}_{GH} & = &
\left( -{z,}_{-}u \right)
\end{array}
     						\label{eq:currents}
\eeq
The one algebraically nonstandard feature of these currents is the
constant term in ${J}_{uC}$ that arises from the explicit time
dependence in the decomposition of the field ${X}^{-}$
in~(\ref{eq:cxdecomp}).  This is avoided if we consider instead the
algebraic relations involving the (also conserved) current
${\tilde{J}}_{uC} \equiv {J}_{uC} - \frac{1}{2\alpha '}$.

Direct calculation of the BRST charge is slightly more involved,
because one must first substitute the fields shifted
by~(\ref{eq:qdelta}) into the integral~(\ref{eq:qcorrfn}),
which gives modified equations of constraint
from the \( {B}_{f}, \, {\hat{B}}_{--}, \, B, \, b \) integrations.
Imposed on the remaining (also shifted) actions by integration of
\( {\hat{X}}^{+}, \, {\hat{g}}^{--}, \, \sqrt{g}, \, c \), these yield
the so-called ``on shell BRST variation''.  The resulting Noether charge
from this variation is related to the currents \( {J}_{u} \) and
\( {J}_{v} \) by
\beq
Q = \int d \sigma
         \left[ u{{J}_{u}}_{C} + v{{J}_{v}}_{C} + \frac{1}{2}
                \left( u{{J}_{u}}_{GH} + v{{J}_{v}}_{GH} \right) \right]
     						\label{eq:qcurrent}
\eeq

We may now check the value of \( {Q}^{2} \) and its relation to the
operator products of the currents that generate the residual symmetry.
Reading from the ghost action and the form of the propagator for
\( {\hat{g}}^{++} \) with \( {\hat{X}}^{-} \) it is possible to show
that the analytic fields have the propagators
\beq
\begin{array}{lcccc}
\left< fh' \right> & = & \left< gj' \right> & = &
                 - \frac{1}{({{\sigma }^{-}}-{{\sigma }^{-}}')} \\
\left< vz' \right> & = & \left< wu' \right> & = &
                   \frac{1}{({{\sigma }^{-}}-{{\sigma }^{-}}')}
\end{array}
     						\label{eq:props}
\eeq
The terms in the operator product expansions for the
currents~(\ref{eq:currents}) singular at short
distance are
\beq
\begin{array}{lcl}
{\tilde{J}}_{uC} {\tilde{J}}_{uC}'  & \sim &
  \frac{28}{2{({{\sigma }^{-}}-{{\sigma }^{-}}')}^{4}}
+ \frac{2{{\tilde{J}}_{uC}}'}{{({{\sigma }^{-}}-{{\sigma }^{-}}')}^{2}}
+ \frac{{{{\tilde{J}}_{uC}',}_{-'}}}{{({{\sigma }^{-}}-{{\sigma }^{-}}')}} \\
{J}_{vC} {J}_{vC}'  & \sim & 0 \\
{\tilde{J}}_{uC} {J}_{vC}'  & \sim &
\frac{{{{J}_{vC}',}_{-'}}}{{({{\sigma }^{-}}-{{\sigma }^{-}}')}} \\
{J}_{uGH} {J}_{uGH}' & \sim &
  \frac{-28}{2{({{\sigma }^{-}}-{{\sigma }^{-}}')}^{4}}
+ \frac{2{J}_{uGH}'}{{({{\sigma }^{-}}-{{\sigma }^{-}}')}^{2}}
+ \frac{{{J}_{uGH}',}_{-'}}{{({{\sigma }^{-}}-{{\sigma }^{-}}')}} \\
{J}_{vGH} {J}_{vGH}' & \sim & 0 \\
{J}_{uGH} {J}_{vGH}' & \sim &
\frac{{{{J}_{vGH}',}_{-'}}}{{({{\sigma }^{-}}-{{\sigma }^{-}}')}}
\end{array}
     						\label{eq:algebra}
\eeq
The functional integral~(\ref{eq:rgcorrfn}) described a decoupled product of
canonical and ghost theories, which are now seen to posess the same
algebra of residual symmetry.  The currents separately generate U(1)
Kac-Moody algebras, with \( {\tilde{J}}_{uC} \) a conformal tensor
of rank $2$ and central charge $28$, \( {J}_{uGH} \) a conformal
tensor of rank $2$ and central charge $-28$, and
\( {J}_{vC} \) and \( {J}_{vGH} \) conformal
scalars.

Using the relation~(\ref{eq:qcurrent}), the operator product
\( {Q}^{2} \) is a
double integral over the positions of two currents on a spatial slice
of the worldsheet, in which only single pole terms from the OPEs of the
currents survive\cite{Tzani}.  A direct calculation of this product
reveals that the pole is proportional to the total central charge of
the theory,
\beq
{Q}^{2} \propto
{C}_{TOT} = {C}_{C} + {C}_{GH} = 0
\eeq
Therefore \( {Q}^{2} = 0 \).  This completes the proof of gauge
invariance of the full quantum theory.

\ch{Comments}

Several things about the previous construction deserve mention, and a
few puzzles remain.  The first notable point
is the number of degrees of freedom of the residual symmetry in
the gauge-fixed functional integral.  In the light-cone
theory with transverse dimensions and no dilaton\cite{Tzani}
the only residual degree of freedom was in one set
of analytic reparametrizations.  The associated ghost current
had central charge $-26$ and the current from the canonical sector had
central charge \( D \).  Thus, the condition \( D = 26 \) was
necessary for the nilpotency of the BRST charge.  In this theory
without transverse dimensions we have replaced
the standard Weyl transformation with~(\ref{eq:pseudoweyl}), and so admitted
another degree of freedom between the light cone and
\( {\hat{g}}^{--} \) ghost equations of motion.  This changes the ghost
central charge from $-26$ to $-28$.

This would have been a disaster for the canonical sector without the
linear dilaton, because there are are no adjustable parameters in the
\( D = 2 \) theory analogous to the number of transverse embedding
dimensions.  The linear dilaton comes to the rescue in a strange way,
though.  Because in the work of Myers\cite{Myers} its gradient was
constrained in terms of the number of transverse dimensions, and in
the work of Tzani\cite{Tzani} they are linked to the nilpotency of the
BRST symmetry, one may have expected the dilaton gradient to appear in
the central charge of this theory.
However, rather than generate a term in the central charge
proportional to \( {\frac{1}{m}}^{2} \) which could be tuned to make
the total charge 28, it introduces another whole degree of freedom
into the solutions of the canonical equations of motion very similar
to that in the ghost sector.  In both sectors there are two conjugate
analytic pairs, the elements of one pair multiplied by a constant
(\( f,h \) and \( u,w \)) and
the elements of the other multiplied by exponentials of
\( \sqrt{2}m\tau \) (\( g,j \) and \( v,z \)).

In the canonical sector the constant pair, which looks like the usual
zero-dilaton solution, contributes \( c = 2 \) and the exponential pair
contributes \( c = 26 \).  In this way the addition of the dilaton
mimics the substitution of the transverse dimensions (which had
\( c = 24 \)) with a conformal field theory with \( c = 26 \).
In the ghost sector the constant pair, which look like the
usual reparametrization ghosts, contribute \( c = -26 \) while the
exponential pair account for the remaining \( c = -2 \).  In all of these
central terms the particular value of \( \frac{1}{m} \) never appears,
though it is still present in the BRST charge and can presumably be
expected to effect the cohomology and hence the physical spectrum.
The important conclusion is that the inclusion of the linear dilaton
{\em at all }, together with this choice for the defining symmetries
of the theory, is enough to guarantee the nilpotency of the BRST
symmetry in the quantum theory.

Actually, the fact that the dilaton gradient is not constrained
by anomaly cancellation is not really a surprise, because in the
total absence of transverse dimensions the previous considerations
do not apply.  After finding the appropriately modified
Lorentz generators, Myers\cite{Myers} was forced to fix the value of
\( {n}^{2} \) to
ensure vanishing of an anomaly in the commutator of the unfixed
lightlike generator with the transverse generators.  This
theory has no transverse directions, so the potentially anomalous
term never appears.  This relates to one of the puzzles.
In that work, the entire physical spectrum was given by
excitations of the transverse oscillators.  Therefore the absence of
transverse embedding dimensions in this theory could be expected to
lead to a trivial physical spectrum.  However, we have found a
nontrivial BRST charge which may well lead to a nontrivial
spectrum\footnote{In our discussions, the whole relation of the
results obtained here to those reported by Myers has been a source
of confusion to the author and to J. Polchinski.}.
More generally, the relation of the continuum theory, in which
physical excitations are expected to be purely transverse, with the
matrix models, which have a nontrivial physical spectrum even with
central charge one, is very confusing.  That relation and its
implications for string field theory are currently
under investigation.

It may yet be that the value of \( \frac{1}{m} \) will be constrained
by something like Lorentz invariance in another way, but the whole
issue of spacetime
Lorentz invariance in this theory is still unclear.  In addition to
explicitly breaking it by introducing the constant vector \( n \), we
have further singled out the same direction by mixing it with the
metric when we defined the shifted variables~(\ref{eq:weylfield}).  The
preservation of Lorentz invariance when it is not manifest is usually
regarded as the central problem in light cone gauge theories, so
defining a useful notion of it here remains an important and central
issue, which I hope also to address in future work.

\ch{Acknowledgement}

I am delighted to thank J. Polchinski for many useful discussions.
I have benefitted throughout this work from his guidance,
both from his ideas and from his unerring sense of direction.
\sppt

\pagebreak

\end{document}